\documentclass[%
 reprint,
 amsmath,amssymb,
 aps,
]{revtex4-1}

\usepackage{graphicx}% Include figure files
\usepackage{dcolumn}% Align table columns on decimal point
\usepackage{bm}% bold math
\usepackage[dvipsnames]{xcolor}
\usepackage{amssymb}
\usepackage{amsmath}

\usepackage[urlcolor=blue, colorlinks=true, bookmarks=false]{hyperref}
\begin{document}

%\preprint{APS/123-QED}

\title{Scale, Quality, and Speed: three key attributes to measure the performance of near-term quantum computers}% Force line breaks with \\

\author{Andrew Wack}
\author{Hanhee Paik}
\author{Ali Javadi-Abhari}
\author{Petar Jurcevic}
\author{Ismael Faro}
\author{Jay M. Gambetta}
\author{Blake R. Johnson}
\affiliation{IBM Quantum, IBM T. J. Watson Research Center, Yorktown Heights, NY 10598}%Lines break automatically or can be forced with \\

%\affiliation{%

\date{\today}% It is always \today, today,

\begin{abstract}
Defining the right metrics to properly represent the performance of a quantum computer is critical to both users and developers of a computing system. In this white paper, we identify three key attributes for quantum computing performance: quality, speed, and scale. Quality and scale are measured by quantum volume and number of qubits, respectively. Using an update to the quantum volume experiments, we propose a speed benchmark that allows the measurement of Circuit Layer Operations Per Second (CLOPS) and identify how both classical and quantum components play a role in improving performance. We prescribe a procedure for measuring CLOPS and use it to characterize the performance of some IBM Quantum systems. 
\end{abstract}

\pacs{Valid PACS appear here}% PACS, the Physics and Astronomy
\maketitle
%\tableofcontents
\section{\label{sec:intro}Introduction}

As we build and deploy increasingly capable quantum computers, it is important to develop benchmarks that track the performance of typical user workloads. Such benchmarks not only aid in identification of the optimal quantum computing system for a particular application, but these metrics also guide performance improvements for system developers. With the ultimate goal of achieving quantum advantage --- where a user can run a quantum program to find a solution faster, cheaper, or more accurately than classical computing alone --- it is important to make progress along the critical factors that drive quantum computing systems toward advantage on meaningful applications. As we enter the era where quantum applications also use substantial classical processing alongside quantum resources, we must be careful to include quantum-classical interactions in the defined benchmarks. Otherwise, our metrics will not be representative of real applications. 

Quantum computing performance is defined by the amount of useful work accomplished by a quantum computer per unit of time. In a quantum computer, the information processing is actualized by quantum circuits containing instructions to manipulate quantum data. Unlike classical computer systems, where instructions are executed directly by a CPU, the Quantum Processing Unit (QPU), which is the combination of the control electronics and quantum memory, is supported by a classical runtime system for converting the circuits into a form consumable by the QPU and then retrieving results for further processing. Performance on actual applications depends on the performance of the complete system, and as such any performance metric must holistically consider all of the components. 

In this white paper, we propose that the performance of a quantum computer is governed by three key factors: scale, quality, and speed. \emph{Scale}, or the number of qubits, determines the size of problem that can be encoded and solved. \emph{Quality} determines the size of quantum circuit that can be faithfully executed. And \emph{speed} is related to the number of primitive circuits that the quantum computing system can execute per unit of time. We introduce a benchmark for measuring speed in section~\ref{sec:speed}. Before diving into the metrics, we first revisit some foundational definitions for clarity.

\section{\label{sec:definitions}Definitions}

\begin{figure*}
\centering
\includegraphics[width=\textwidth]{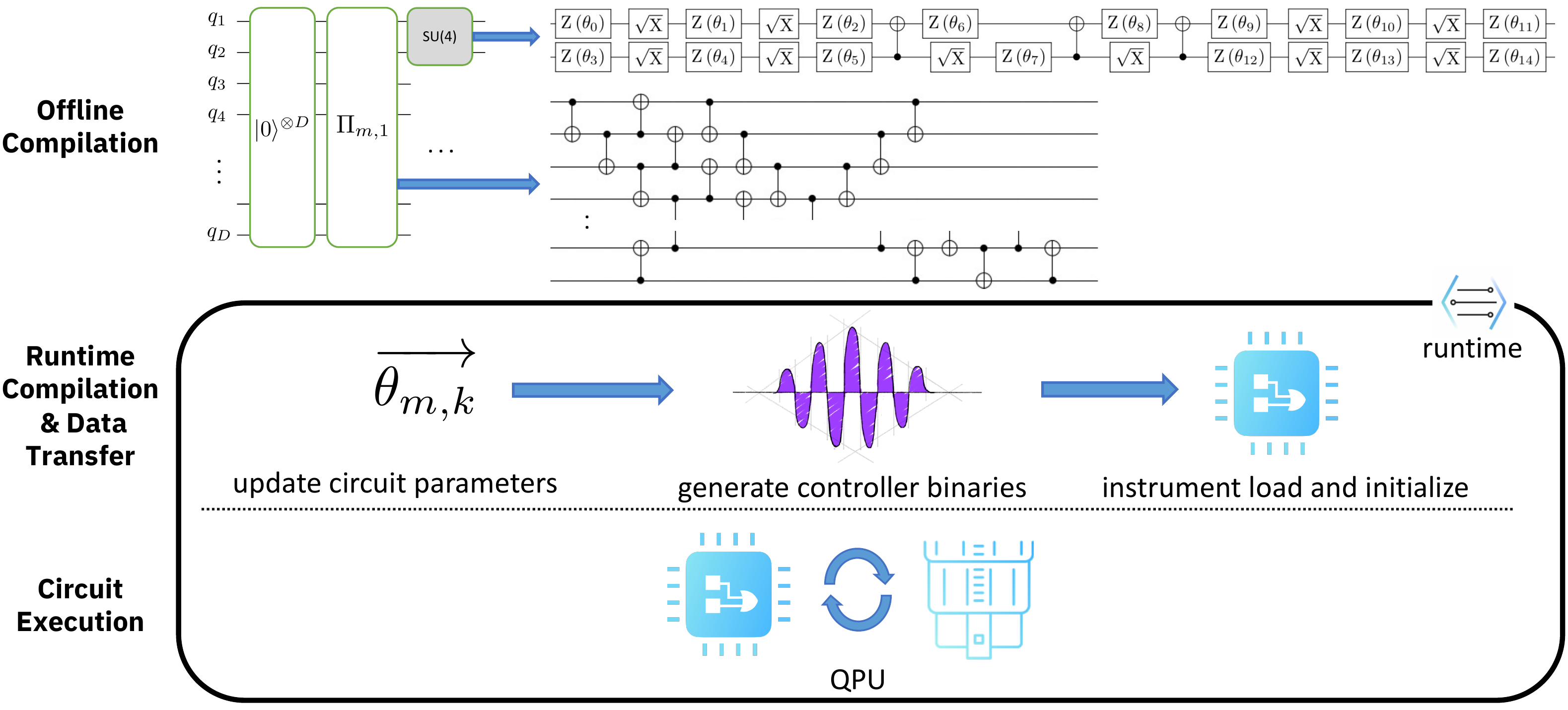}
\caption{Runtime architecture and phases of compilation. The circuit pattern of a Quantum Volume benchmark is shown, as well as its offline compilation. Circuit parameters in the Circuit Layer Operations per Second benchmark are updated during runtime, making the metric heavily dependent on the runtime architecture and runtime compilation.}
\label{fig:compilation-phases}
\end{figure*}
\subsection{\label{sec:circuit-def}Quantum Circuits and Programs}

A \emph{quantum circuit} is a computational routine consisting of an ordered sequence of quantum operations including gates, measurements, and resets on quantum data (qubits) and concurrent real-time classical computation~\cite{Corcoles21a, Cross21a}. Data flows between the quantum operations and the real-time classical compute so that the classical compute can incorporate measurement results and the quantum operations may be conditioned upon or parameterized by data from the real-time classical compute. Here \emph{real-time} means within the coherence time of the qubits. We exclude from the quantum circuit definition \emph{near-time} computations occurring on time scales longer than the coherence of the quantum computation. Extended quantum circuits may be completely described by the OpenQASM 3 language~\cite{Cross21a}.

A \emph{quantum program} expresses a larger concept of a task or algorithm that executes or samples from multiple quantum circuits within the context of a larger classical computer program. Variational algorithms are examples of quantum programs that execute circuits within a classical optimization loop. For these workloads, system performance increases substantially with a runtime architecture that hosts the classical computation in a \emph{near-time} context with low-latency access to quantum hardware. Quantum-classical interactions occur via the classical program requesting execution of one or more quantum circuits. The communication time for these requests plays a large role in system performance, thus we deem it essential to capture this time in defining a speed metric.

\subsection{Compiling quantum programs: offline \& runtime compilation}

A key part of executing quantum programs is the compilation of quantum circuits that arise from them. Figure~\ref{fig:compilation-phases} shows our envisioned phases of compilation and how they interact with the runtime. Many quantum programs, such as variational ones, give rise to quantum circuits whose outlines are known, but whose parameters are runtime-dependent.

In the Quantum Volume benchmark~\cite{Cross19a} the circuit patterns are composed of layers of qubit permutation and random 2-qubit unitaries. These can be compiled and optimized to a great degree offline. Permutations capture data motion, a key requirement of large-scale quantum computing. These may be compiled to SWAP networks, CNOT networks, teleportation, shuttling instructions, etc. Random unitaries capture average circuits and how they may explore the Hilbert space. These may be compiled using a variety of methods to the target gate set~\cite{Jurcevic20a}. The quality of the circuits will be dictated to a great extent by how well this compiler performs.

The benchmark proposed here provisions some parameters that will only be available during runtime (e.g. those that depend on previous runs). The runtime compiler is thus responsible for binding such parameters, generating new binaries, and feeding them to the control electronics of the QPU. Repeated interactions between quantum-classical compute make this critical for speed, which is what the benchmark aims to capture. We have thus designed the benchmark so as to separate these phases, and measure speed as it pertains to the interactive use of the QPU. Of course the separation is not exact: knowing the parameters or latest calibration data can help the compiler improve the circuits even more, through approximation, noise adaptivity, etc.~\cite{Cross19a,Murali20}. If this information is used, then it will be included in the benchmark measurement as well.

\section{\label{sec:metrics}Performance metrics}
\begin{figure}
\centering
\includegraphics[width=.5\textwidth]{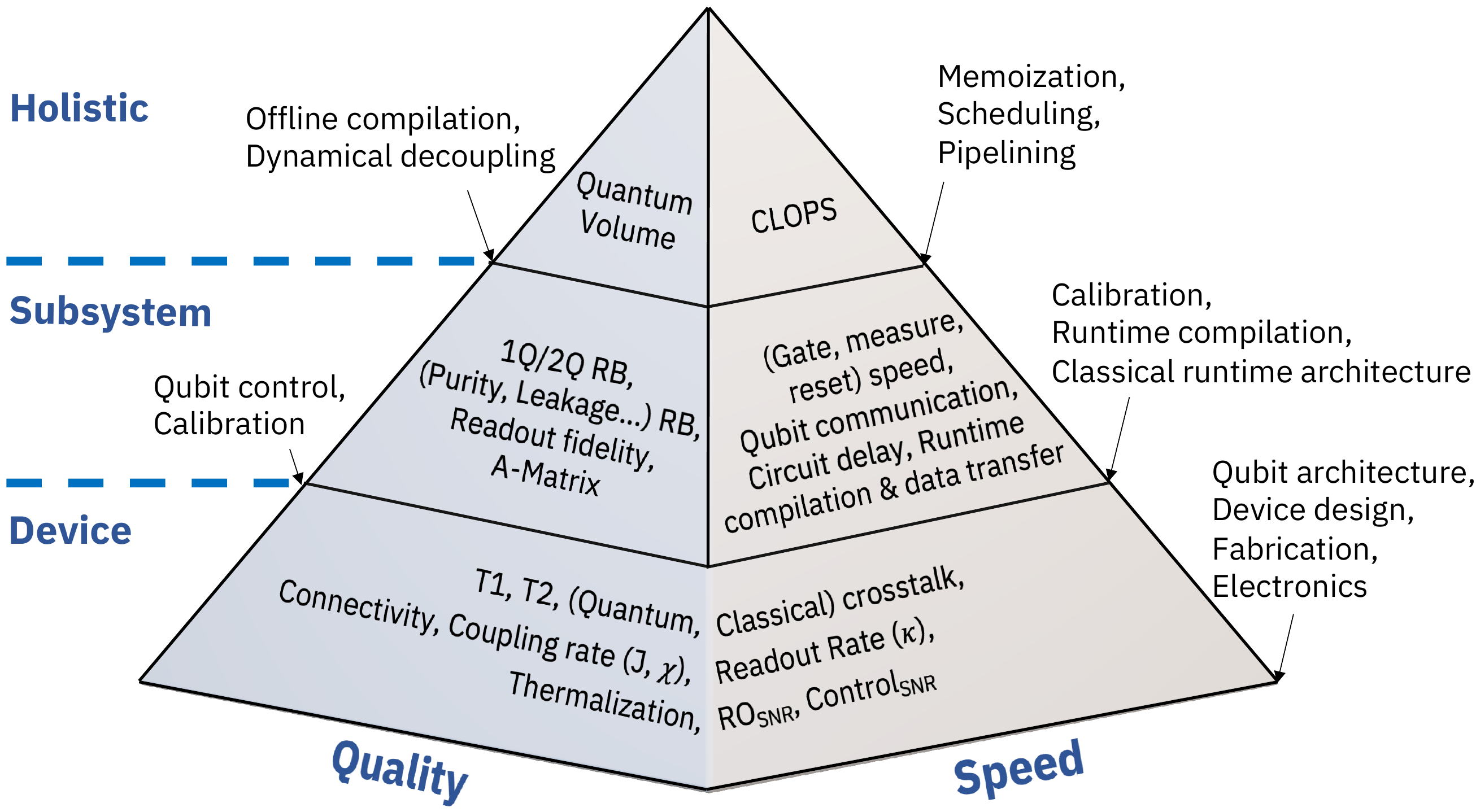}
\caption{Benchmarking pyramid showing how quality and speed can be benchmarked. Higher-level benchmarks capture more complexity but less specificity. There may be tradeoffs between the two faces of the pyramid.}
\label{fig:pyramid}
\end{figure}

In this section we describe the three metrics in more detail. Before we begin, it is worth emphasizing that many design choices are made in building a quantum computer, all of which can affect the quality, speed, and scale. A useful way of thinking about this is a benchmarking pyramid~\cite{pyramid}, where different levels of complexity are captured at different tiers. Device level parameters provide more complexity, but do not give an accurate picture of the overall performance. Holistic benchmarks capture the many different ways that system parameters can interact and influence the overall performance, at the cost of less specificity. In Figure~\ref{fig:pyramid}, we enhance this picture by including another face to the pyramid: speed. This figure shows some of the ``ingredients'' that influence quality and speed, which we capture holistically using Quantum Volume and Circuit Layer Operations Per Second metrics.

\subsection{\label{sec:scale}Number of qubits}
The number of qubits determines the amount of information that can be encoded in a quantum computer for computation, which caps the size of solvable problems. For example, in chemistry simulations, the number of qubits sets the size of the basis set that represents each electron wavefunction in a molecule and therefore the size of the molecules that can be simulated. Since fault-tolerant computation requires very large number of qubits, scale is a key metric in the development of quantum computers. Number of qubits can also be used as a resource to improve the other two metrics of quality and speed. For example, auxiliary qubits can often be used to reduce the depth of circuits and increase their fidelity~\cite{hoyer05}. Extra qubits can also be used in multiprogramming of QPUs to increase their circuit processing speed~\cite{das19}.

For most quantum computing platforms, increasing the number of qubits --- the scalability --- relies heavily on the available materials and fabrication technologies developed from the semiconducting industry.  Superconducting, semiconducting, ion traps and photonics qubit platforms all leverage 3D integration technology and multi-layer fabrication process from CMOS packaging for scalable on-chip wiring solutions, building traps and building photonic waveguides.  While all quantum hardware platforms have challenges in scaling, superconducting qubits are making fast progress to scale beyond 100 qubits. 

The challenge lies rather in developing key technologies that make scaling possible while maintaining quantum coherence on the processor. Technology development for quantum hardware takes a significant amount of investment and time in hard tech, and this is the reason we must continue investing and developing technologies for quantum computing hardware.

\begin{figure*}
\centering
\includegraphics[width=\textwidth]{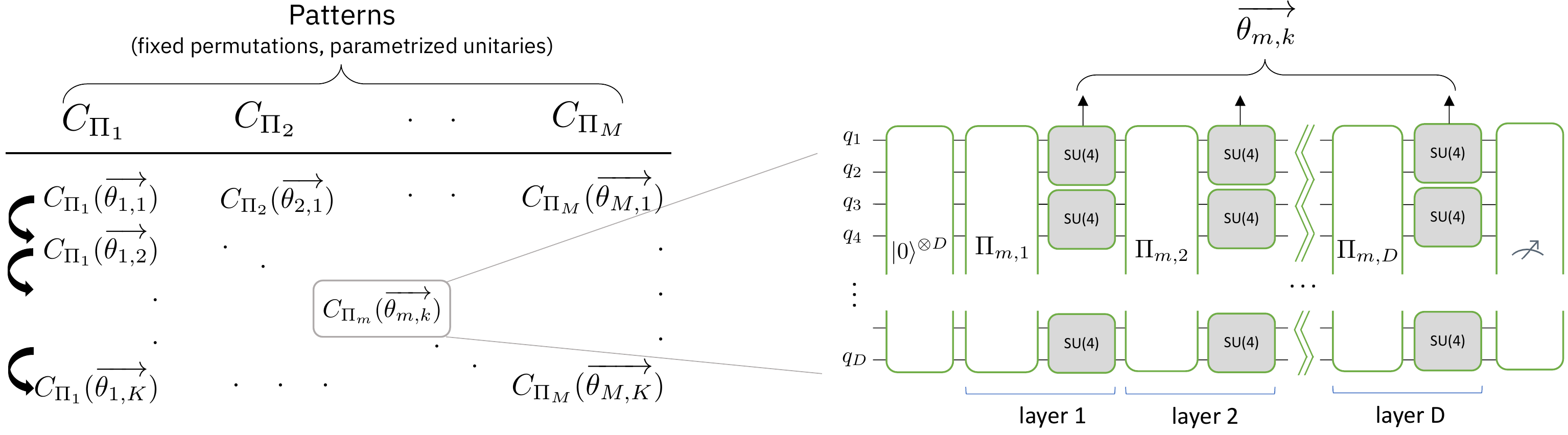}
\caption{\textbf{Matrix of circuits used for CLOPS benchmark} There are $M=100$ independent templates of QV circuits, with $D$ layers of SU(4)s, where each SU(4) in the circuit is fully parameterized. The parameters for each circuit are updated $K=10$ times. The parameters $\theta_{m,k}$ depend on the output from circuit using parameters $\theta_{m,k-1}$ }
\label{fig:QPU-speed-circuits}
\end{figure*}

\subsection{\label{sec:quality}Quantum volume}
Quantum volume (QV)~\cite{Cross19a} indicates how faithfully a quantum circuit can be implemented in a quantum computing system. We define a \emph{QV layer} as one layer of permutation among qubits and one layer of pair-wise random SU(4) 2-qubit unitary gates, as shown in Fig.~\ref{fig:QPU-speed-circuits}. The QV is defined by the \emph{width} or number of QV layers of the largest random square circuit (with width equal to the number of layers) that a quantum processor can successfully run. Note that when a QV circuit is compiled to the native gate set of a particular QPU, the circuit depth of the compiled circuit will typically be much larger than the number of QV layers as the abstract permutations and SU(4) unitaries may be each decomposed into multiple native gates. QV measurement starts with executing a square circuit of width $N$, and then compares the measurement results from the heavy output states (the states with probabilities higher than median of probabilities of all output states) with the ideal results from simulation. The largest $N$-qubit, square circuit that can run successfully to produce more than 2/3 of heavy outputs, determines the quantum volume on a quantum computing system, given by $2^N$.   

Quantum volume is sensitive to coherence, gate fidelity, and measurement fidelity which are hardware properties of a quantum processor. Quantum volume is also influenced by connectivity and compilers which can make circuits efficient to minimize the effect of decoherence~\cite{Jurcevic20a}.  Quantum volume is a holistic metric because it cannot be improved by just improving one aspect of the system, but rather requires all parts of the system to be improved in a synergistic manner. Quantum volume has been adopted widely by research and industry, and has been reported for several ion trap and superconducting quantum computers.

\subsection{\label{sec:speed}Circuit Layer Operations per Second Benchmark}

Circuit Layer Operations per Second (CLOPS) is a measure correlated with how many QV circuits a QPU can execute per unit of time. That simple statement hides a wealth of choice about the possible circuit families and the execution context. Here, we pursue a holistic speed benchmark of a \emph{typical} application. In order to faithfully model real-world use, we deem it essential to capture interaction time with the run-time environment that invokes the circuits. This attempts to avoid a pitfall seen in some synthetic benchmarks that characterize classical systems by their instruction clock rate without considering the effects of data transfers between CPU, cache and main memory. With more fragile quantum data we don't have the luxury to persist quantum data across multiple invocations, so data transfer plays an even more prominent role. To capture this element, we suggest to measure multiple executions of parameterized circuits where the choice of parameters is deferred until run time. This mimics the scenario found in algorithms such as the variational quantum eigensolver (VQE) or quantum kernel alignment, where the ability of a quantum system to efficiently handle parameterized circuits is key to performance of these algorithms.

Measuring execution speed of \emph{typical} applications requires either a corpus of representative circuits or a choice of circuit family that somehow captures an ``average'' circuit. While acknowledging the difficulty in the latter, we argue that QV circuits are at least representative of random circuits while simultaneously allowing for a rigorous notion of quality. This last point ensures that the benchmarked circuits operate in a regime where the QPU is producing meaningful results. A natural extension of the metric proposed here would consider speed for a variety of other circuit families~\cite{lubinski2021}.

\begin{figure*}
\begin{center}
\newcolumntype{T}[1]{%
    >{\centering\arraybackslash\hspace{0pt}}p{#1}}%
\begin{tabular}{ T{0.15\textwidth} | T{0.1\textwidth}| T{0.1\textwidth} | T{0.10\textwidth} | T{0.1\textwidth} | T{0.10\textwidth}}
& \multicolumn{4}{c|}{Attributes} & \\
Device & Qubits & QV & Layers & Shots & CLOPS\\ 
 \hline
 ibmq\_bogota & 5 & 32 & 5 &  100 & 1419 \\
 \hline
ibmq\_toronto & 27 & 32 & 5 &  100 & 951\\ 
 \hline
 ibmq\_brooklyn & 65 & 32 & 5 & 100 & 753
\end{tabular}
\caption{\textbf{CLOPS results}}
\label{fig:QPU-speed-results}
\end{center}
\end{figure*}

We formally define CLOPS as the number QV layers executed per second using a set of parameterized QV circuits, where each QV circuit has $D = \log_2\text{QV}$ layers. Circuit execution time includes updating parameters to the circuit, submitting the job to the QPU, executing on the QPU and sending results back to be processed. CLOPS is then calculated as the total number of QV layers executed divided by the total execution time.

The CLOPS benchmark consists of 100 parameterized template circuits (denoted $C_{\Pi_m}$ for $ 0 \leq m < 100$) of the same type as the model circuits used when measuring the quantum volume of the system except that each of the SU(4) random unitaries are left fully parameterized. In other words, we choose and fix the random permutation layers in 100 QV circuits while leaving the SU(4)s on adjacent qubit pairs parameterized. Each circuit template $C_{\Pi_m}$ will be executed 10 times with 10 choices of random parameters ($\overrightarrow{\theta_{m,k}}, 0 \leq m < 100, 0 \leq k < 10$).  These parameters are applied to the circuit template to generate the final circuit that is then run on the system without any further parameter updates. Each of these instantiated circuits are denoted as $C_{\Pi_m}(\overrightarrow{\theta_{m,k}})$. See figure~\ref{fig:QPU-speed-circuits}. Each of these circuits is executed with 100 shots, which is an attempt to balance the benchmark between just measuring setup time for execution against the number of shots typically required to estimate an observable with reasonable variance from the output. The benchmark procedure is as follows:

\begin{enumerate}
    \item The 100 parameterized circuits, $C_{\Pi_m}$, are generated and may be compiled to a parameterized circuit in the target machine's native gate set
    \item Time is started
    \item The parameters for the initial circuits $\overrightarrow{\theta_{m,0}}$ are generated using a suitable pseudo-random number generator (PRNG), and used to run circuits $C_{\Pi_m}(\overrightarrow{\theta_{m,0}})$
    \item For $C_{\Pi_m}(\overrightarrow{\theta_{m,k}})$ where $k>0$ the output of $C_{\Pi_m}(\overrightarrow{\theta_{m,k-1}})$ is used to seed the PRNG to generate parameters $\overrightarrow{\theta_{m,k}}$. Thus circuit $C_{\Pi_m}(\overrightarrow{\theta_{m,k}})$ may not run until $C_{\Pi_m}(\overrightarrow{\theta_{m,k-1}})$ has completed and returned its results.
    \item Circuits are executed using 100 shots and the same qubits, gate length, inter-circuit delays, etc. that were used when establishing the QV of the device
    \item Outside of the constraints listed in item 4 and 5, circuits may be executed in any order or combination that makes the best use of the system resources while not changing the QV quality.
    \item When all circuits have been run and results received, time is stopped.
\end{enumerate}

\newpage
CLOPS can then be calculated as:

\begin{equation*}
    \frac{M \times K \times S \times D}{\textrm{time\_taken}}
\end{equation*}
\begin{equation*}
    \begin{aligned}
        \textrm{where:}\\
        M & = \textrm{number of templates} = 100\\
        K & = \textrm{number of parameter updates} = 10\\
        S & = \textrm{number of shots} = 100\\
        D & = \textrm{number of QV layers} = \log_2\textrm{QV}\\
    \end{aligned}
\end{equation*}

%\begin{equation}
%        \frac{\sum_{m=0}^{m<100} d(C_{\Pi_m}) \times \textrm{num\_parameter\_updates} \times \textrm{shots}}{\textrm{time\_taken}}
%\end{equation}

\begin{figure*}
\centering
\newcolumntype{T}[1]{%
    >{\centering\arraybackslash\hspace{0pt}}p{#1}}%
\begin{tabular}{ T{0.15\textwidth} | T{0.1\textwidth}| T{0.1\textwidth} | T{0.1\textwidth} | T{0.1\textwidth} T{0.1\textwidth} }
 & & \multicolumn{3}{c}{Time breakdown}  \\
 &  Total & Circuit &  Circuit & \multicolumn{2}{c}{Run-time compilation \&} \\ 
Device & time & execution & delay & \multicolumn{2}{c}{data transfer}\\ 
& & & & & \\[-\normalbaselineskip]
 \hline
 ibmq\_bogota & 352.2 & 2.5 & 25.0  &  \multicolumn{2}{c}{324.7}  \\
 \hline
ibmq\_toronto & 525.7 & 2.4 & 25.0 & \multicolumn{2}{c}{498.4} \\ 
 \hline
 ibmq\_brooklyn & 663.6 & 2.0 & 25.0 & \multicolumn{2}{c}{636.6}

\end{tabular}
\caption{\textbf{CLOPS time breakdown} All times are in seconds}
\label{fig:QPU-speed-breakdown}
\end{figure*}

The CLOPS benchmark is designed to allow the system to leverage all of the quantum resources on a device to run a collection of circuits as fast as possible, as well as stress all parts of the execution pipeline. This includes data transfer of circuits and results, run-time compilation (lowering basis-gate level circuits to control electronics instructions), latencies in loading control electronics, initialization of control electronics, gate times, measurement times, reset time of qubits, delays between circuits, processing results as well as parameterized updates. The generation of random parameters from a seed constructed from the shots of the previous execution simulates the parameter updates in iterative quantum algorithms.

Including all of these parameters in the benchmark ensures that all aspects of the system are included to generate a meaningful comparison between systems. Physical qubit architectures may effect the repetition time, gate times, reset times, and measurement times and can vary significantly across technologies. For example, the repetition rate and the gate rate of superconducting qubits~\cite{Corcoles21a} can be orders of magnitude faster than the ones of ion trap qubits~\cite{Pino21a} which significantly impacts the CLOPS. Similarly, software components such as run-time compilation, orchestration of the control electronics, etc. are all aspects that are necessary in current architectures to run already ``compiled'' programs for the QPU and have considerable impact on overall performance.  Finally the CLOPS is also impacted by how efficiently circuits can be delivered to the system for execution and results returned to the user-space application. 

\section{\label{sec:level1}Measurement of performance metrics of IBM quantum computing systems}

We have run the CLOPS benchmark on several of our systems.  We chose systems that had the same quantum volume, but varied in size to highlight the current differences in their performance. The results are shown in figure \ref{fig:QPU-speed-results}. The systems range in size from 5 to 65 qubits, each with quantum volume 32. By choosing machines with the same QV, the characteristics of the benchmark were the same for each machine (e.g. size of circuits, typical gate depth in the native gate set, etc).  Additional each of these machines have the same \emph{rep\_delay} (delay between shots) of 250 microseconds.   Despite all of these similarities, we see differences in performance, with the largest machine performing the slowest, at a CLOPS of 753 layers per second, compared to 1419 for the 5 qubit device. Given the similarities we might expect that the numbers to be nearly identical, but the benchmark reveals real world differences in performance that the user experiences. 

To better understand those differences, we have broken down the time to execute the benchmark on each device into five areas: 
\begin{enumerate}
    \item Time actually spent running the circuit on the device: \textit{circuit execution}
    \item Delay time between each shot of each circuit on the device: \textit{circuit delay}
    \item Time spent on preparing the circuits to actually run on device (parameter upddates, run-time compilation, waveform generation) as well as data transfers (circuit submission to the backend, instrument initialization, instrument load, return of results to user) : \textit{run-time compilation and data transfer}
\end{enumerate}

The time breakdowns for each device are shown in figure~\ref{fig:QPU-speed-breakdown}. This breakdown highlights the need for a holistic benchmark of speed as we can see from the time to execute the circuits themselves, the machine that would appear to be fastest, \textit{ibmq\_brooklyn} is in fact the slowest. This is because other factors dominate, \emph{circuit delay} being one order of magnitude larger than gate time, and  most notably \emph{run-time compilation and data transfer} which two orders of magnitude larger, limiting the overall utilization. This shows the clear value of the benchmark to allow us to find the real barriers to improved speed. 

The first of these large consumers, \emph{circuit delay}, represents the idle time between circuits on the device, and we can see that this is the same for all of the benchmarked systems, which is expected as the default delay is the same on all systems and we ran the same number of circuits and shots. While today this does not dominate the benchmark, as we drive down the \emph{run-time compilation and data transfer} it will have a considerable effect on the CLOPS.  Also for applications that require larger shot counts, this factor becomes more dominant as the rest of the overheads do not scale with shot count. Because of these factors, reducing this delay time is a constant area of research, and as quantum devices improve we expect that this delay time can be dramatically shortened. 

The second area, \emph{run-time compilation and data transfer}, shows effects of increasing time with machine size. This happens for several reasons. The first is that the larger machines require larger complexes of control electronics to send signals to the device, and this requires more time to initialize and load. As we move forward, changes in our software stack will improve these characteristics, reducing initialization effort as well as amount of data needed to be loaded. The second piece is the time to compile the circuits into the instructions needed for the control electronics. As we move to OpenQASM3, we are building a new high performance compiler to do this final lowering step, which we expect to have a large impact on performance. Similarly, support for parameterized updates throughout the stack will help reduce the run-time compilation requirements as well as reducing amount of data that needs to be moved.

%The \emph{communication} time is an example of where architecture changes can have a dramatic impact on performance. While \emph{communication} time is currently a relatively small portion of the total time, historically this was not the case. Our benchmark was run taking advantage of the Qiskit Runtime environment, which tightly couples the user environment with the backend for faster delivery of circuits to the backend and results back to the user. If this were run through the standard cloud interface, it would be a dominant component of the overall time, anywhere from 4 to 30 times higher, reflecting the longer delays in the upper stack and typical queue wait times, all of which are bypassed by the Qiskit Runtime.

\section{\label{sec:conclusion}Summary}

Performance benchmarks have always been difficult to properly engineer for classical computer systems, and quantum systems add both result quality and interaction with classical systems into the equation. We have shown that low level, single dimension benchmarks do not properly express the performance that user's see from the system. Instead it is necessary to create holistic benchmarks that capture all of the components that will translate to performance on real world applications but not be overly cumbersome to execute. We have defined a CLOPS benchmark that captures many of the necessary aspects for running user applications with good performance, and provided examples of using the benchmark to find current bottlenecks in the system.

\bibliography{references}

\begin{thebibliography}{10}%
\makeatletter
\providecommand \@ifxundefined [1]{%
 \@ifx{#1\undefined}
}%
\providecommand \@ifnum [1]{%
 \ifnum #1\expandafter \@firstoftwo
 \else \expandafter \@secondoftwo
 \fi
}%
\providecommand \@ifx [1]{%
 \ifx #1\expandafter \@firstoftwo
 \else \expandafter \@secondoftwo
 \fi
}%
\providecommand \natexlab [1]{#1}%
\providecommand \enquote  [1]{``#1''}%
\providecommand \bibnamefont  [1]{#1}%
\providecommand \bibfnamefont [1]{#1}%
\providecommand \citenamefont [1]{#1}%
\providecommand \href@noop [0]{\@secondoftwo}%
\providecommand \href [0]{\begingroup \@sanitize@url \@href}%
\providecommand \@href[1]{\@@startlink{#1}\@@href}%
\providecommand \@@href[1]{\endgroup#1\@@endlink}%
\providecommand \@sanitize@url [0]{\catcode `\\12\catcode `\$12\catcode
  `\&12\catcode `\#12\catcode `\^12\catcode `\_12\catcode `\%12\relax}%
\providecommand \@@startlink[1]{}%
\providecommand \@@endlink[0]{}%
\providecommand \url  [0]{\begingroup\@sanitize@url \@url }%
\providecommand \@url [1]{\endgroup\@href {#1}{\urlprefix }}%
\providecommand \urlprefix  [0]{URL }%
\providecommand \Eprint [0]{\href }%
\providecommand \doibase [0]{http://dx.doi.org/}%
\providecommand \selectlanguage [0]{\@gobble}%
\providecommand \bibinfo  [0]{\@secondoftwo}%
\providecommand \bibfield  [0]{\@secondoftwo}%
\providecommand \translation [1]{[#1]}%
\providecommand \BibitemOpen [0]{}%
\providecommand \bibitemStop [0]{}%
\providecommand \bibitemNoStop [0]{.\EOS\space}%
\providecommand \EOS [0]{\spacefactor3000\relax}%
\providecommand \BibitemShut  [1]{\csname bibitem#1\endcsname}%
\let\auto@bib@innerbib\@empty
%</preamble>
\bibitem [{\citenamefont {C\'orcoles}\ \emph {et~al.}(2021)\citenamefont
  {C\'orcoles}, \citenamefont {Takita}, \citenamefont {Inoue}, \citenamefont
  {Lekuch}, \citenamefont {Minev}, \citenamefont {Chow},\ and\ \citenamefont
  {Gambetta}}]{Corcoles21a}%
  \BibitemOpen
  \bibfield  {author} {\bibinfo {author} {\bibfnamefont {A.~D.}\ \bibnamefont
  {C\'orcoles}}, \bibinfo {author} {\bibfnamefont {M.}~\bibnamefont {Takita}},
  \bibinfo {author} {\bibfnamefont {K.}~\bibnamefont {Inoue}}, \bibinfo
  {author} {\bibfnamefont {S.}~\bibnamefont {Lekuch}}, \bibinfo {author}
  {\bibfnamefont {Z.~K.}\ \bibnamefont {Minev}}, \bibinfo {author}
  {\bibfnamefont {J.~M.}\ \bibnamefont {Chow}}, \ and\ \bibinfo {author}
  {\bibfnamefont {J.~M.}\ \bibnamefont {Gambetta}},\ }\href {\doibase
  10.1103/PhysRevLett.127.100501} {\bibfield  {journal} {\bibinfo  {journal}
  {Phys. Rev. Lett.}\ }\textbf {\bibinfo {volume} {127}},\ \bibinfo {pages}
  {100501} (\bibinfo {year} {2021})}\BibitemShut {NoStop}%
\bibitem [{\citenamefont {Cross}\ \emph {et~al.}(2021)\citenamefont {Cross},
  \citenamefont {Javadi-Abhari}, \citenamefont {Alexander}, \citenamefont
  {de~Beaudrap}, \citenamefont {Bishop}, \citenamefont {Heidel}, \citenamefont
  {Ryan}, \citenamefont {Smolin}, \citenamefont {Gambetta},\ and\ \citenamefont
  {Johnson}}]{Cross21a}%
  \BibitemOpen
  \bibfield  {author} {\bibinfo {author} {\bibfnamefont {A.~W.}\ \bibnamefont
  {Cross}}, \bibinfo {author} {\bibfnamefont {A.}~\bibnamefont
  {Javadi-Abhari}}, \bibinfo {author} {\bibfnamefont {T.}~\bibnamefont
  {Alexander}}, \bibinfo {author} {\bibfnamefont {N.}~\bibnamefont
  {de~Beaudrap}}, \bibinfo {author} {\bibfnamefont {L.~S.}\ \bibnamefont
  {Bishop}}, \bibinfo {author} {\bibfnamefont {S.}~\bibnamefont {Heidel}},
  \bibinfo {author} {\bibfnamefont {C.~A.}\ \bibnamefont {Ryan}}, \bibinfo
  {author} {\bibfnamefont {J.}~\bibnamefont {Smolin}}, \bibinfo {author}
  {\bibfnamefont {J.~M.}\ \bibnamefont {Gambetta}}, \ and\ \bibinfo {author}
  {\bibfnamefont {B.~R.}\ \bibnamefont {Johnson}},\ }\href@noop {} {\enquote
  {\bibinfo {title} {Openqasm 3: A broader and deeper quantum assembly
  language},}\ } (\bibinfo {year} {2021}),\ \Eprint
  {http://arxiv.org/abs/2104.14722} {arXiv:2104.14722 [quant-ph]} \BibitemShut
  {NoStop}%
\bibitem [{\citenamefont {Cross}\ \emph {et~al.}(2019)\citenamefont {Cross},
  \citenamefont {Bishop}, \citenamefont {Sheldon}, \citenamefont {Nation},\
  and\ \citenamefont {Gambetta}}]{Cross19a}%
  \BibitemOpen
  \bibfield  {author} {\bibinfo {author} {\bibfnamefont {A.~W.}\ \bibnamefont
  {Cross}}, \bibinfo {author} {\bibfnamefont {L.~S.}\ \bibnamefont {Bishop}},
  \bibinfo {author} {\bibfnamefont {S.}~\bibnamefont {Sheldon}}, \bibinfo
  {author} {\bibfnamefont {P.~D.}\ \bibnamefont {Nation}}, \ and\ \bibinfo
  {author} {\bibfnamefont {J.~M.}\ \bibnamefont {Gambetta}},\ }\href {\doibase
  10.1103/PhysRevA.100.032328} {\bibfield  {journal} {\bibinfo  {journal}
  {Phys. Rev. A}\ }\textbf {\bibinfo {volume} {100}},\ \bibinfo {pages}
  {032328} (\bibinfo {year} {2019})}\BibitemShut {NoStop}%
\bibitem [{\citenamefont {Jurcevic}\ \emph {et~al.}(2020)\citenamefont
  {Jurcevic}, \citenamefont {Javadi-Abhari}, \citenamefont {Bishop},
  \citenamefont {Lauer}, \citenamefont {Bogorin}, \citenamefont {Brink},
  \citenamefont {Capelluto}, \citenamefont {Günlük}, \citenamefont {Itoko},
  \citenamefont {Kanazawa}, \citenamefont {Kandala}, \citenamefont {Keefe},
  \citenamefont {Krsulich}, \citenamefont {Landers}, \citenamefont
  {Lewandowski}, \citenamefont {McClure}, \citenamefont {Nannicini},
  \citenamefont {Narasgond}, \citenamefont {Nayfeh}, \citenamefont {Pritchett},
  \citenamefont {Rothwell}, \citenamefont {Srinivasan}, \citenamefont
  {Sundaresan}, \citenamefont {Wang}, \citenamefont {Wei}, \citenamefont
  {Wood}, \citenamefont {Yau}, \citenamefont {Zhang}, \citenamefont {Dial},
  \citenamefont {Chow},\ and\ \citenamefont {Gambetta}}]{Jurcevic20a}%
  \BibitemOpen
  \bibfield  {author} {\bibinfo {author} {\bibfnamefont {P.}~\bibnamefont
  {Jurcevic}}, \bibinfo {author} {\bibfnamefont {A.}~\bibnamefont
  {Javadi-Abhari}}, \bibinfo {author} {\bibfnamefont {L.~S.}\ \bibnamefont
  {Bishop}}, \bibinfo {author} {\bibfnamefont {I.}~\bibnamefont {Lauer}},
  \bibinfo {author} {\bibfnamefont {D.~F.}\ \bibnamefont {Bogorin}}, \bibinfo
  {author} {\bibfnamefont {M.}~\bibnamefont {Brink}}, \bibinfo {author}
  {\bibfnamefont {L.}~\bibnamefont {Capelluto}}, \bibinfo {author}
  {\bibfnamefont {O.}~\bibnamefont {Günlük}}, \bibinfo {author}
  {\bibfnamefont {T.}~\bibnamefont {Itoko}}, \bibinfo {author} {\bibfnamefont
  {N.}~\bibnamefont {Kanazawa}}, \bibinfo {author} {\bibfnamefont
  {A.}~\bibnamefont {Kandala}}, \bibinfo {author} {\bibfnamefont {G.~A.}\
  \bibnamefont {Keefe}}, \bibinfo {author} {\bibfnamefont {K.}~\bibnamefont
  {Krsulich}}, \bibinfo {author} {\bibfnamefont {W.}~\bibnamefont {Landers}},
  \bibinfo {author} {\bibfnamefont {E.~P.}\ \bibnamefont {Lewandowski}},
  \bibinfo {author} {\bibfnamefont {D.~T.}\ \bibnamefont {McClure}}, \bibinfo
  {author} {\bibfnamefont {G.}~\bibnamefont {Nannicini}}, \bibinfo {author}
  {\bibfnamefont {A.}~\bibnamefont {Narasgond}}, \bibinfo {author}
  {\bibfnamefont {H.~M.}\ \bibnamefont {Nayfeh}}, \bibinfo {author}
  {\bibfnamefont {E.}~\bibnamefont {Pritchett}}, \bibinfo {author}
  {\bibfnamefont {M.~B.}\ \bibnamefont {Rothwell}}, \bibinfo {author}
  {\bibfnamefont {S.}~\bibnamefont {Srinivasan}}, \bibinfo {author}
  {\bibfnamefont {N.}~\bibnamefont {Sundaresan}}, \bibinfo {author}
  {\bibfnamefont {C.}~\bibnamefont {Wang}}, \bibinfo {author} {\bibfnamefont
  {K.~X.}\ \bibnamefont {Wei}}, \bibinfo {author} {\bibfnamefont {C.~J.}\
  \bibnamefont {Wood}}, \bibinfo {author} {\bibfnamefont {J.-B.}\ \bibnamefont
  {Yau}}, \bibinfo {author} {\bibfnamefont {E.~J.}\ \bibnamefont {Zhang}},
  \bibinfo {author} {\bibfnamefont {O.~E.}\ \bibnamefont {Dial}}, \bibinfo
  {author} {\bibfnamefont {J.~M.}\ \bibnamefont {Chow}}, \ and\ \bibinfo
  {author} {\bibfnamefont {J.~M.}\ \bibnamefont {Gambetta}},\ }\href@noop {}
  {\enquote {\bibinfo {title} {Demonstration of quantum volume 64 on a
  superconducting quantum computing system},}\ } (\bibinfo {year} {2020}),\
  \Eprint {http://arxiv.org/abs/2008.08571} {arXiv:2008.08571 [quant-ph]}
  \BibitemShut {NoStop}%
\bibitem [{\citenamefont {Murali}\ \emph {et~al.}(2020)\citenamefont {Murali},
  \citenamefont {McKay}, \citenamefont {Martonosi},\ and\ \citenamefont
  {Javadi-Abhari}}]{Murali20}%
  \BibitemOpen
  \bibfield  {author} {\bibinfo {author} {\bibfnamefont {P.}~\bibnamefont
  {Murali}}, \bibinfo {author} {\bibfnamefont {D.~C.}\ \bibnamefont {McKay}},
  \bibinfo {author} {\bibfnamefont {M.}~\bibnamefont {Martonosi}}, \ and\
  \bibinfo {author} {\bibfnamefont {A.}~\bibnamefont {Javadi-Abhari}},\ }in\
  \href@noop {} {\emph {\bibinfo {booktitle} {Proceedings of the Twenty-Fifth
  International Conference on Architectural Support for Programming Languages
  and Operating Systems}}}\ (\bibinfo {year} {2020})\ pp.\ \bibinfo {pages}
  {1001--1016}\BibitemShut {NoStop}%
\bibitem [{\citenamefont {McKay}\ and\ \citenamefont {Merkel}(2021)}]{pyramid}%
  \BibitemOpen
  \bibfield  {author} {\bibinfo {author} {\bibfnamefont {D.}~\bibnamefont
  {McKay}}\ and\ \bibinfo {author} {\bibfnamefont {S.}~\bibnamefont {Merkel}},\
  }\href@noop {} {\bibfield  {journal} {\bibinfo  {journal} {Private
  communication}\ } (\bibinfo {year} {2021})}\BibitemShut {NoStop}%
\bibitem [{\citenamefont {H{\o}yer}\ and\ \citenamefont
  {{\v{S}}palek}(2005)}]{hoyer05}%
  \BibitemOpen
  \bibfield  {author} {\bibinfo {author} {\bibfnamefont {P.}~\bibnamefont
  {H{\o}yer}}\ and\ \bibinfo {author} {\bibfnamefont {R.}~\bibnamefont
  {{\v{S}}palek}},\ }\href@noop {} {\bibfield  {journal} {\bibinfo  {journal}
  {Theory of computing}\ }\textbf {\bibinfo {volume} {1}},\ \bibinfo {pages}
  {81} (\bibinfo {year} {2005})}\BibitemShut {NoStop}%
\bibitem [{\citenamefont {Das}\ \emph {et~al.}(2019)\citenamefont {Das},
  \citenamefont {Tannu}, \citenamefont {Nair},\ and\ \citenamefont
  {Qureshi}}]{das19}%
  \BibitemOpen
  \bibfield  {author} {\bibinfo {author} {\bibfnamefont {P.}~\bibnamefont
  {Das}}, \bibinfo {author} {\bibfnamefont {S.~S.}\ \bibnamefont {Tannu}},
  \bibinfo {author} {\bibfnamefont {P.~J.}\ \bibnamefont {Nair}}, \ and\
  \bibinfo {author} {\bibfnamefont {M.}~\bibnamefont {Qureshi}},\ }in\
  \href@noop {} {\emph {\bibinfo {booktitle} {Proceedings of the 52nd Annual
  IEEE/ACM International Symposium on Microarchitecture}}}\ (\bibinfo {year}
  {2019})\ pp.\ \bibinfo {pages} {291--303}\BibitemShut {NoStop}%
\bibitem [{\citenamefont {Lubinski}\ \emph {et~al.}(2021)\citenamefont
  {Lubinski}, \citenamefont {Johri}, \citenamefont {Varosy}, \citenamefont
  {Coleman}, \citenamefont {Zhao}, \citenamefont {Necaise}, \citenamefont
  {Baldwin}, \citenamefont {Mayer},\ and\ \citenamefont
  {Proctor}}]{lubinski2021}%
  \BibitemOpen
  \bibfield  {author} {\bibinfo {author} {\bibfnamefont {T.}~\bibnamefont
  {Lubinski}}, \bibinfo {author} {\bibfnamefont {S.}~\bibnamefont {Johri}},
  \bibinfo {author} {\bibfnamefont {P.}~\bibnamefont {Varosy}}, \bibinfo
  {author} {\bibfnamefont {J.}~\bibnamefont {Coleman}}, \bibinfo {author}
  {\bibfnamefont {L.}~\bibnamefont {Zhao}}, \bibinfo {author} {\bibfnamefont
  {J.}~\bibnamefont {Necaise}}, \bibinfo {author} {\bibfnamefont {C.~H.}\
  \bibnamefont {Baldwin}}, \bibinfo {author} {\bibfnamefont {K.}~\bibnamefont
  {Mayer}}, \ and\ \bibinfo {author} {\bibfnamefont {T.}~\bibnamefont
  {Proctor}},\ }\href@noop {} {\enquote {\bibinfo {title} {Application-oriented
  performance benchmarks for quantum computing},}\ } (\bibinfo {year} {2021}),\
  \Eprint {http://arxiv.org/abs/2110.03137} {arXiv:2110.03137 [quant-ph]}
  \BibitemShut {NoStop}%
\bibitem [{\citenamefont {Pino}\ \emph {et~al.}(2021)\citenamefont {Pino},
  \citenamefont {Dreiling}, \citenamefont {Figgatt}, \citenamefont {Gaebler},
  \citenamefont {Moses}, \citenamefont {Allman}, \citenamefont {Baldwin},
  \citenamefont {Foss-Feig}, \citenamefont {Hayes}, \citenamefont {Mayer},
  \citenamefont {Ryan-Anderson},\ and\ \citenamefont {Neyenhuis}}]{Pino21a}%
  \BibitemOpen
  \bibfield  {author} {\bibinfo {author} {\bibfnamefont {J.~M.}\ \bibnamefont
  {Pino}}, \bibinfo {author} {\bibfnamefont {J.~M.}\ \bibnamefont {Dreiling}},
  \bibinfo {author} {\bibfnamefont {C.}~\bibnamefont {Figgatt}}, \bibinfo
  {author} {\bibfnamefont {J.~P.}\ \bibnamefont {Gaebler}}, \bibinfo {author}
  {\bibfnamefont {S.~A.}\ \bibnamefont {Moses}}, \bibinfo {author}
  {\bibfnamefont {M.~S.}\ \bibnamefont {Allman}}, \bibinfo {author}
  {\bibfnamefont {C.~H.}\ \bibnamefont {Baldwin}}, \bibinfo {author}
  {\bibfnamefont {M.}~\bibnamefont {Foss-Feig}}, \bibinfo {author}
  {\bibfnamefont {D.}~\bibnamefont {Hayes}}, \bibinfo {author} {\bibfnamefont
  {K.}~\bibnamefont {Mayer}}, \bibinfo {author} {\bibfnamefont
  {C.}~\bibnamefont {Ryan-Anderson}}, \ and\ \bibinfo {author} {\bibfnamefont
  {B.}~\bibnamefont {Neyenhuis}},\ }\href@noop {} {\bibfield  {journal}
  {\bibinfo  {journal} {Nature}\ }\textbf {\bibinfo {volume} {592}},\ \bibinfo
  {pages} {209} (\bibinfo {year} {2021})}\BibitemShut {NoStop}%
\end{thebibliography}%

\clearpage

\section{\label{sec:appendix}Appendix: Depth-1 circuit performance}

While the QV layers provide a means to compare across different quantum computer architectures, it is also useful to explore the speed of particular systems relative to their own basis gate set. We can do this with some simple additional instrumentation on the CLOPS benchmark. This provides a means to extrapolate from the CLOPS results to how it might affect other algorithms when run on the same system. We first define how to measure circuit depth in this context and then show some results for our systems.

\subsection{Circuit depth}
Circuit depth is an important parameter in performance metrics. With current quantum computing systems, depth plays a key role in result quality. For speed metrics we need a consistent way to count operations performed by the QPU. 

There are several considerations we need to take when calculating depth.  First, there may be gates used that the machine cannot natively execute. For those gates we assume that they are decomposed into the native gate set of the machine before depth is calculated. Second, unlike classical systems, operations in algorithms on qubits are assumed to operate with full parallelism. We can think of each of the gates as occupying boxes on parallel timelines for each qubit in the circuit. Single qubit gates are boxes that span a single unit of time and act only on a single qubit. Two qubit gates result in a box that spans a single unit of time but encompasses two qubits.  Since two qubit gates need to operate synchronously on both qubits, the box acts as a implicit barrier and we cannot place the box in the time sequence until both qubits are ready to execute the gate.  Once we have all boxes placed, we can then look for the longest sequence of boxes across all of the qubits to determine the depth of the circuit.

Formally, the following rules are then used to calculate the depth $d$  for each qubit in the circuit
\begin{itemize}
    \item depth of a 1 qubit gate of non-zero duration from the native gate set is 1 
    \item depth of a measurement is 1 
    \item depth of reset is 1 
    \item depth of a 2 qubit gate from the native gate set is 1 and is preceded by an implicit barrier on the same 2 qubits and the barrier rule applies 
    \item depth of a barrier is 0, but synchronizes the depth across all applied qubits to the current max depth of those qubits at the barrier
\end{itemize}
The depth $d$ for circuit $C$ with qubits $Q_0\dots Q_n$ is then : 
\begin{equation}
d(C) =
max[d(Q_0), d(Q_1),\dots, d(Q_n)]
\end{equation}
where $Q_0...Q_n$ are all of the qubits in the circuit

The above definitions allows us to define a useful concept of ``depth-1 circuits'' as a primitive circuit. A depth-1 circuit is any circuit that can be executed across a set of qubits in a single unit of time from the depth definition above. This allows us to talk about any circuit of depth $d$ as an ordered sequence of $d$ depth-1 circuits, and the performance of a quantum computing system in terms of its speed of execution of these depth-1 circuits.

\subsection{Measurement of depth-1 circuit performance}

Figure~\ref{fig:depth-1-results} Updates the CLOPS table to include the depth-1 circuits per second executed. We instrumented the CLOPS code so that when the benchmark was run, the average depth of each of the random 100 circuit templates is reported.  All machines reported the same depth value at around 99, a reflection that each used a similar qubit topology to run the benchmark, a straight line of 5 qubits. From this depth value, knowing the number of circuits and shots run as well as the total time we can calculate the depth-1 circuits per second. Because all of these values are the same for these machines, this results in a linear scaling from the CLOPS values to the depth-1 circuits per second values. 
\begin{figure*}[h!]
\begin{center}
\newcolumntype{T}[1]{%
    >{\centering\arraybackslash\hspace{0pt}}p{#1}}%
\begin{tabular}{ T{0.15\textwidth} | T{0.1\textwidth}| T{0.1\textwidth} | T{0.10\textwidth} | T{0.1\textwidth} | T{0.10\textwidth} | T{0.10\textwidth}}
& \multicolumn{4}{c|}{Attributes} &  &\\
 &  &  &  &  &  & depth-1 circ \\ 
Device & Qubits & QV & Layers & Shots & CLOPS & per second\\ 
 \hline
 ibmq\_bogota & 5 & 32 & 5 &  100 & 1419 & 28355 \\
 \hline
ibmq\_toronto & 27 & 32 & 5 &  100 & 951 & 18837\\ 
 \hline
 ibmq\_brooklyn & 65 & 32 & 5 & 100 & 753 & 15041
\end{tabular}
\caption{\textbf{CLOPS results with depth-1 circuits per second}}
\label{fig:depth-1-results}
\end{center}
\end{figure*}
\end{document}